\documentclass[english]{article}

\usepackage{geometry} 
\usepackage{graphicx}
\usepackage{amssymb}
\usepackage{epstopdf}
\usepackage{babel}
\usepackage{inputenc}
\usepackage{epsfig}

\begin{document}

%\begin{flushright} 
%Draft, \today \\ 
%\end{flushright}

\vspace{0.8cm}

\begin{center}
{\Large {\bf Hadron-nuclei collisions at high energies}} 
\vspace{0.8cm}

\normalsize{
G. Giacomelli 

{\small
Physics Department of the University of Bologna and INFN Sezione di Bologna, I-40127 Bologna, Italy}
giacomelli@bo.infn.it
}

\vspace{1cm}

{\large \bf In honour of Dumitru B. Ion}

\end{center}

\vspace{0.5cm}

{\bf Abstract.} {\normalsize 
A brief historical review is made of the hadron-hadron ($hh$) total cross section and hadron-nucleus absorption cross section measurements, made mainly at high energy proton synchrotrons. Then I shall discuss low $p_{t}$ processes, including diffraction processes and fragmentation of nuclei in nucleus-nucleus collisions. Nucleus-nucleus collisions at higher energy colliders are then considered, mainly in the context of the search for the gluon quark plasma. Conclusions and a short discussion on perspectives follow.}

\section{Introduction}

In the second half of the 20th century several high-quality secondary beams of charged particles became available at proton synchrotrons of increasing energy: the PS and SPS at CERN, the AGS at BNL, the 70 GeV PS at IHEP Serpukhov, and the PS at Fermilab. The secondary charged hadron beams contained the six stable or quasi stable charged particles, $\pi^{\pm}$, $K^{\pm}$, $p$, $\bar{p}$, and very small backgrounds (mainly muons and electrons). Several experiments on hadron production \cite{Giac}, on total hadron-hadron ($hh$) cross sections \cite{Giac, cool, allaby, carroll2} and hadron-nucleus absorption cross sections \cite{binon} were performed, using the transmission method in ``good geometry". These experiments were refined, relatively simple and were conducted by a relatively small number of physicists and engineers from few different Institutions.

    In some high-intensity beams the production of the charged hadrons was studied in more detail \cite{baker}, and several particle searches were made, which also lead to the study of $\bar{d}$, $\bar{t}$, and $\overline{He}^{3}$ production \cite{bez}.
 
    In order to reach higher energies it became necessary to build hadron-hadron colliders. The first was the ISR $pp$ and $\bar{p}p$ collider at CERN, which allowed to reach a maximum c.m. energy of 63 GeV; then the $S\bar{p}pS$ collider at CERN allowed $\bar{p}p$ collisions up to 600 and 900 GeV c.m. energies;  the Fermilab Tevatron $\bar{p}p$ collider allowed c.m. energies up to 1.8 and recently 2.0 TeV. Finally the RHIC collider at BNL allows $pp$ and nucleus-nucleus collisions at c.m. energies around 200 GeV/nucleon. 
    
    Some of the experiments performed at the colliders were still relatively simple  dedicated experiments, like those for total $hh$ cross sections and for $hh$ elastic scattering measurements \cite{amaldi, bozzo, amos, avila}, and single arm spectrometers for the study of low $p_{t}$ inelastic collisions \cite{antinucci}. But then followed elaborated general purpose detectors \cite{brea, gg3}: the new ones have all a central detector, an electromagnetic calorimeter, a hadron calorimeter and a muon detector, and the experiments have a very large number of electronic channels and need hundreds of physicists and engineers \cite{abe}. Thus also the sociology of the experiments changed considerably \cite{ggi}. Experiments at the RHIC collider use a variety of detectors, most of which are refined and complex detectors with many electronic channels \cite{bnl}.
    
   The next future accelerator will be the Large Hadron Collider (LHC) at CERN, which will allow $pp$ collisions up to c.m. energies of 14 TeV and nucleus-nucleus collisions up to c.m. energies of $\sqrt{s_{NN}}$=5.5 TeV/nucleon. Two general purpose detectors (ATLAS and CMS) are designed to investigate the largest range of physics, while one experiment (ALICE), focuses on the search for the Quark-Gluon Plasma (QGP) \cite{link}. These experiments use large, elaborated detectors with an incredibly large number of electronic channels and very sophisticated and complex computing needs, hardware and software; each experiment was made and will be operated by thousands of physicists and engineers. Other specialized detectors are designed for b-physics (LHCb), total cross section (TOTEM) and forward physics (LHCf) \cite{link}. It is expected that LHC will open up a completely new and unexplored energy region.
   
   Higher energies were and still are obtained only with cosmic rays \cite{yodh}; and also cosmic ray experiments are becoming very large, like the Auger experiment \cite{auger}.
   
   In this paper we shall mainly concentrate on the analysis of hadron-nucleus collisions at high energies, on the high energy low $p_{t}$ parameters, some features of the low $p_{t}$ inelastic collisions and briefly on the searches for the Quark Gluon Plasma.

\begin{figure}[t]
\begin{center}
{\centering\resizebox*{!}{8.4cm}{\includegraphics{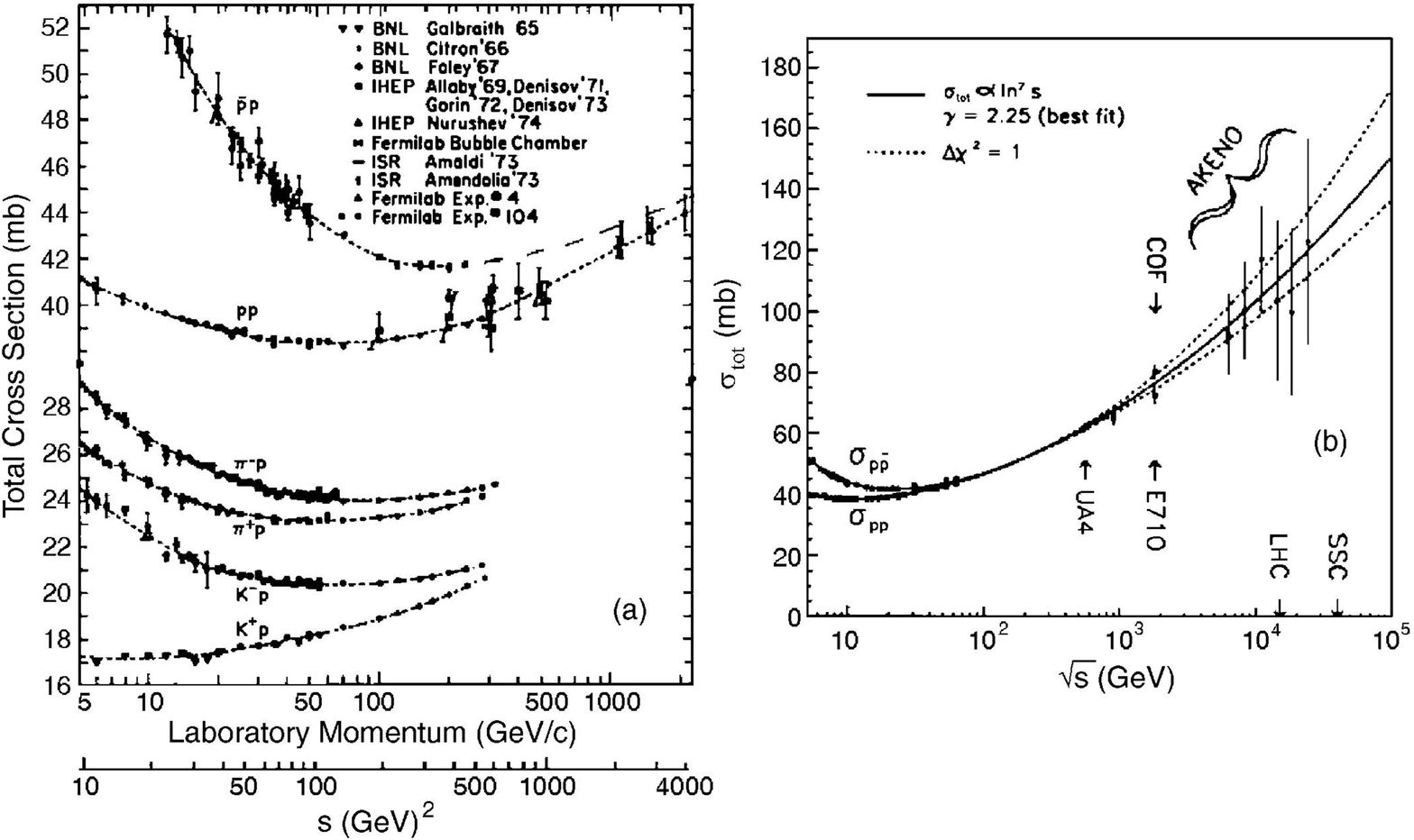}}}
\vspace{0.4cm}
\begin{quote} 
\caption{\small Compilation of total cross-sections (a) for high-energy $hh$ scattering and (b) for higher
energy $\bar{p}p$ and $pp$, including cosmic ray measurements; the solid line is a Regge pole fit; the uncertainty in the energy trend is indicated by dashed lines.}
\label {fig:fig1}
\end{quote}
\end{center}
 \end{figure}

\section{Hadron-hadron total cross sections}

At fixed target proton syncrotron accelerators (at BNL, IHEP and Fermilab) the total cross sections of charged hadrons were measured with the transmission method in good geometry, with relative precisions smaller than 1$\%$ and systematic scale errors of 1-2$\%$ \cite{cool, gg3}. 

   The measurements of the total cross sections at the $\bar{p}p$ and $pp$ colliders required the development of new experimental techniques: the scattering of particles was measured at very small angles, with detectors positioned  in re-entrant containers (``Roman pots") located very close to the circulating beams and far away from the interaction point. The combinations of statistical and systematic uncertainties are $\geq$10$\%$, with the exception of the CERN ISR, where the luminosity was measured accurately by the Van der Meer method of displacing vertically the beams \cite{amaldi, amos, gg3}.
   
   Fig. \ref{fig:fig1}a, b summarize the present status of high energy $hh$ total cross sections: above the resonance region the total $hh$ cross sections decrease, reach a minimum and then increase with increasing energy    
(the $K$+$p$  total cross section was already increasing at IHEP-Serpukhov energies \cite{allaby}). The difference between the $\bar{x}p$ and $xp$ total cross sections decreases with increasing energy.

    At the highest energies there are only cosmic ray (CR) data \cite{yodh}; the very large CR experiments will improve the experimental situation at these very large energies \cite{auger}.\\

\section{ Inelastic low $p_{t}$ processes}

\begin{figure}[!h]
\begin{center}
{\centering\resizebox*{!}{4.4 cm}{\includegraphics{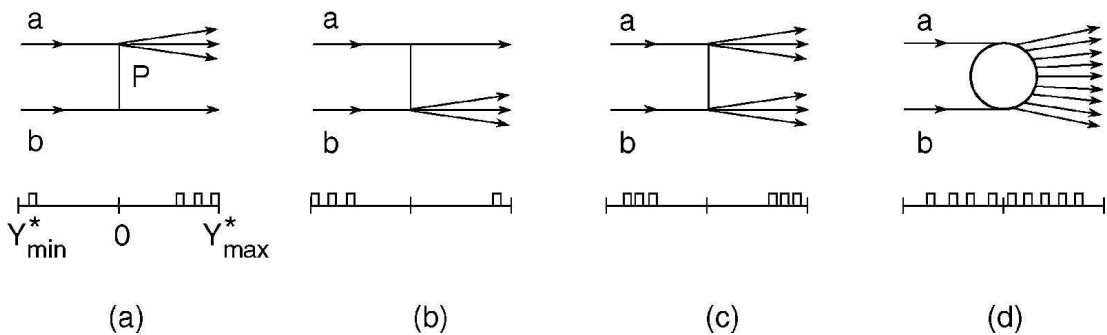}}}     
 %\vspace{0.4cm}
\begin{quote} 
\caption{\small Pictorial description of inelastic, small $p_{t}$ processes with characteristic rapidity distributions.
P indicates Pomeron exchange. (a) Fragmentation of the beam particle a; (b) fragmentation
of the target particle b; (c) double fragmentation of a and b; (d) inelastic non diffractive collisions.}
\label {fig:fig2}
\end{quote}
\end{center}
 \end{figure}

Fig. \ref{fig:fig2} gives a pictorial representation of the dominant inelastic processes in $hh$ collisions at low $p_{t}$: single and double diffraction dissociation and inelastic processes, which concentrate in the forward direction; this  is also the most important direction for cosmic ray data. The total $\bar{p}p$ cross section may be written as
                                                                                                                                                     
\begin{equation}
\sigma_{tot}= \sigma_{el} + \sigma_{inel} = \sigma_{el} + \sigma_{sd} + \bar{\sigma}_{sd} + \sigma_{dd} + \sigma_{nd}                                                                                                                                            
\end{equation}\\
where $\sigma_{el}$ is the elastic cross section, $\sigma_{sd}$ is the single diffractive cross section for the incoming proton, $\bar{\sigma}_{sd}$ is the single diffractive cross section for the incoming antiproton ($\bar{\sigma}_{sd}$ $\simeq$ $\sigma_{sd}$), $\sigma_{dd}$ is the double diffractive cross section and $\sigma_{nd}$ is the non diffractive part of the of the inelastic cross section. $\sigma_{sd}$ and $\sigma_{dd}$ are not  measured very precisely at high energies; most of the non diffractive cross section concerns hadrons emitted with low transverse momenta ($low$ $p_{t}$ $physics$) with properties which change slowly with c.m. energy ($\ln{s}$ $physics$) 
\cite{gg3}.     
                                                                                                                                                                                                                                                                                                                                                                                                                    

   A small part of the non diffractive cross section is due to central collisions  between the two colliding particles  and give rise to high $p_{t}$ jets of particles emitted at large angles ($large$ $p_{t}$ $physics$); this contribution increases with increasing energies and eventually becomes dominant.
   
\begin{figure}[!h]
\begin{center}
\vspace{1cm}
{\centering\resizebox*{!}{7.2 cm}{\includegraphics{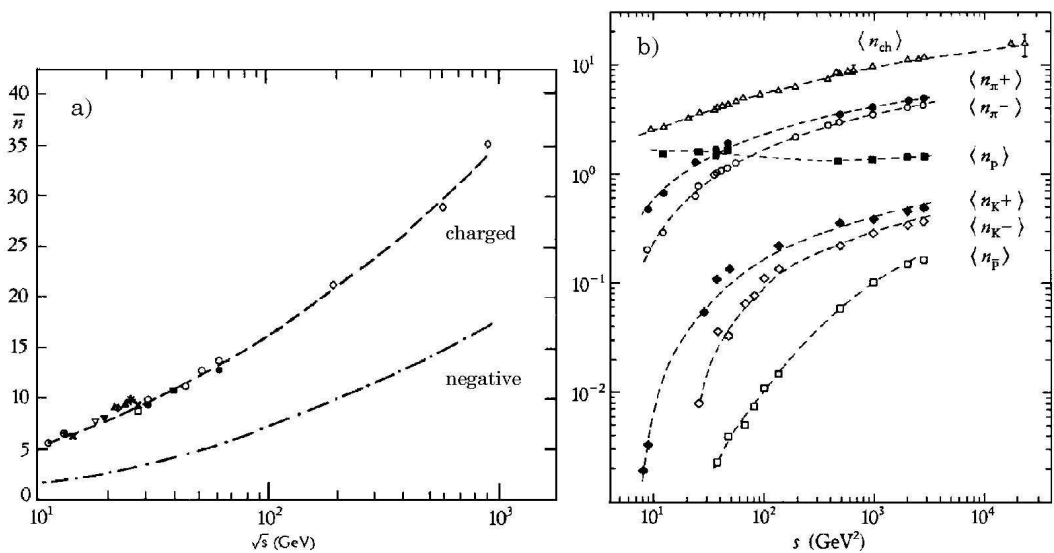}}}
%\vspace{-1cm}
\begin{quote} 
\caption{\small (a) Average charged multiplicities $\bar{n}$ for $\bar{p}p$ and $pp$ collisions $vs$. $E_{cm}$ =$\sqrt{s}$. (b) Average
number of $\pi^{\pm}$, $K^{\pm}$, $p$ and $\bar{p}$ produced in $pp$ collisions at the CERN ISR for c. m. energies up to 63 GeV.}
\label {fig:fig3}
\vspace{-0.5cm}
\end{quote}
\end{center}
 \end{figure}

   The average number of charged hadrons, $<n_{ch}>$, produced in high energy collisions increases with increasing c.m. energy $\sqrt{s}$, as shown in Fig. \ref{fig:fig3}a for the charged multiplicities in $\bar{p}p$ and $pp$ collisions \cite{antinucci, brea}. The data may be fitted to a power law dependence in $\ln{s}$ of the type \cite{binon}:
   
\begin{equation}
<n> = A + B \ln s + C \ln^{2} s \simeq 3.6 - 0.45 \ln s + 0.2 \ln^{2} s
\end{equation}\\
In $\bar{p}p$ collision at $\sqrt{s} =$ 1.8 TeV are produced on average about 40 charged particles and 20 neutral particles in each collision. Fig. \ref{fig:fig3}b shows that most of the produced particles are pions, followed by kaons; it is also visible the so called $leading$ $effect$, which is connected with the emission of relatively many high energy protons in the incoming proton direction (antiprotons in the incoming direction of antiprotons) \cite{antinucci, abe}. The computations of charged multiplicities at higher  energies are based on Monte Carlo methods, which have considerable uncertainties \cite{mor}. Also the measurements are not easy. It is instead easier to measure and calculate the multiplicities from quarks, and gluons jets at large $p_{t}$ \cite{daga}. It is interesting to recall that gluon jets yield larger charged multiplicities than quark jets.  

   At the LHC, for $pp$ collisions at $\sqrt{s} =$ 14 TeV, one expects the production of 70-90 charged particles per collision \cite{mor}.
   
   The average $p_{t}$ of the produced particles increases slowly with $\sqrt{s}$: it was $\sim$0.36 GeV/c in 20$<\sqrt{s}<$100 GeV and it increased to $\sim$0.46 GeV/c at $\sqrt{s} =$ 1.8 TeV. The simplest interpretation of these features is in terms of thermodynamic models \cite{gg3}.
   
\begin{figure}[!t]
\begin{center}
\vspace{-0.4cm}
{\centering\resizebox*{!}{6 cm}{\includegraphics[angle=270]{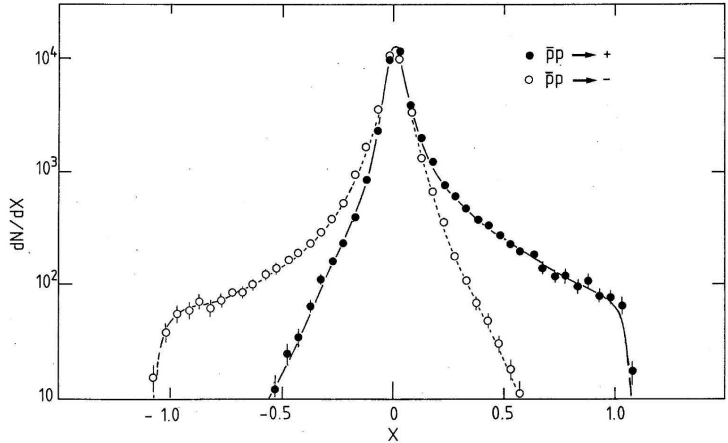}}}
%\vspace{0.6cm}
\begin{quote} 
\caption{\small $x$-distributions of positive and negative particles produced $\bar{p}p$ interactions at $\sqrt{s}$=53 GeV \cite{brea}. Positive values of $x$ correspond to the direction of the incoming protons. The slight differences between leading particles for $x>0$ and $x<0$ are due to the acceptance of the apparatus. The lines are only meant to guide the eye.}
\label {fig:fig4}
\vspace{-0.5cm}
\end{quote}
\end{center}
 \end{figure}

   It may be stressed that some of the main qualitative features of particle production at high energies and low $p_{t}$ are easily observed in the Feynman $x$ distribution ($x=p_{l}/p_{l\,max}$ in the c.m. system). Fig. \ref{fig:fig4} shows the production of positive and negative particles in $\bar{p}p$ interactions at $\sqrt{s}$=53 GeV plotted versus $x$: for $x>0$ (the direction of the incoming protons) the distribution for positive particles contains two components, fragmentation products for $x>0.3$ and centrally produced particles for $x<0.3$. The distribution for negative particles on the other hand has only one component, which is typical of centrally produced particles, i.e. with a large concentration of particles at $x=0$, rapid fall off at larger $x$, and with very few particles for $x>0.3$ \cite{brea, gg3}. In the region for $x<0$ the negative particles are leading particles, following the charge of the incident antiproton, while positive particles are the produced ones. For $pp$ collisions the situation is symmetric around $x=0$ and the distributions show leading effects for the positive particles only.  
   
\section{Absorption cross sections in hadron-nuclei collisions}
\begin{figure}[!h]
\begin{center}
\vspace{0.5cm}
{\centering\resizebox*{!}{7.2 cm}{\includegraphics{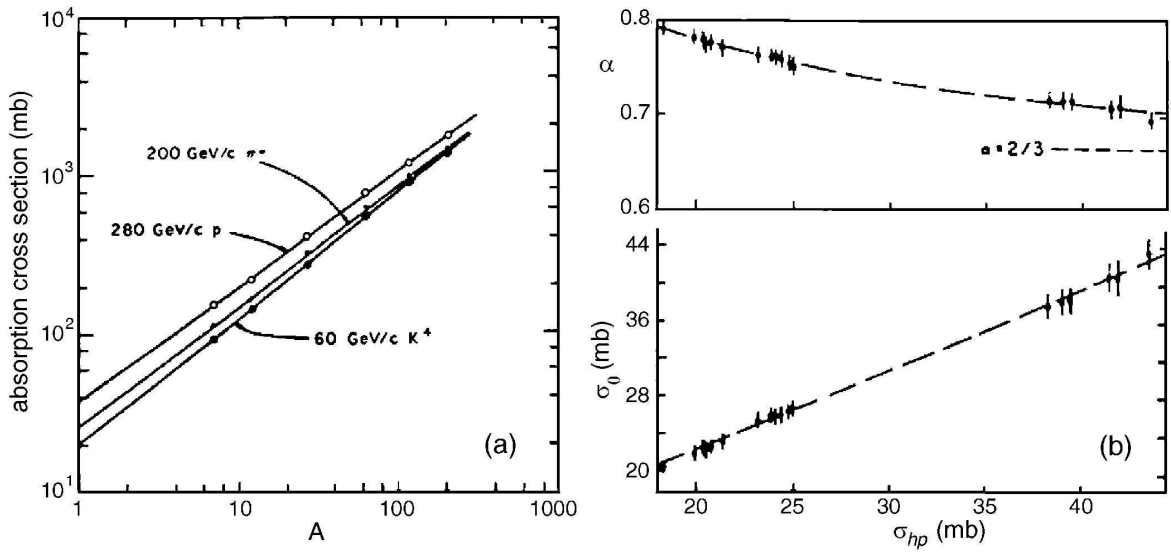}}}
\vspace{-0.4cm}
\begin{quote} 
\caption{\small (a) Absorption cross-sections versus atomic weight A; the solid lines are fits to eq. (3). (b) The parameters $\sigma_{0}$ and $\alpha $ versus the corresponding $hp$ total cross-section ($\sigma_{hp}$).}
\label {fig:fig4b}
\vspace{-0.8cm}
\end{quote}
\end{center}
 \end{figure}

As byproducts of $hh$ total cross section measurements, the absorption cross sections of charged pions, charged kaons, protons and antiprotons were measured on several target nuclei, for example Li, C, Al, Cu, Sn and Pb \cite{binon}. The energy dependence of these cross sections tends to follow that of the $hh$ cross sections, but are somewhat slower. 
   The data at each energy were fitted to the simple expression
                                                                                                                                                          
\begin{equation}
\sigma_{abs}(A) = \sigma_{0} A^{\alpha}
\end{equation}                                                                                                                                                            
where A is the atomic weight of the target nucleus. Examples of the fits are shown in Fig. \ref{fig:fig4b}: notice the good fits to Eq. 3 and that, in the lab momentum range 60-280 GeV/c; the parameter $\sigma_{0}$ increases with energy ( with the exception of antiprotons); the parameter $\alpha$ decreases with increasing energy, with a tendency to go towards the value 2/3 for large values of $\sigma_{hp}$, as would be expected for an opaque nucleus \cite{binon, gg3}. The relations between $hh$ and h-nuclei cross sections are in most cases given in the context of the Glauber theory \cite{glauber}: inelastic scattering is treated in the shadowed single collision approximation at small momentum transfer, multiple collision approximation at large momentum transfer.

\section{Fragmentation of nuclei in nucleus-nucleus collisions} 
\begin{figure}[t!]
%\begin{center}
\centering
% {\centering\resizebox*{7.7cm}{5.1cm}{\includegraphics{hadron-hadron1.jpg}}}
 \hspace{-0.9cm}
 {\centering\resizebox*{15.5cm}{4.5cm}{\includegraphics{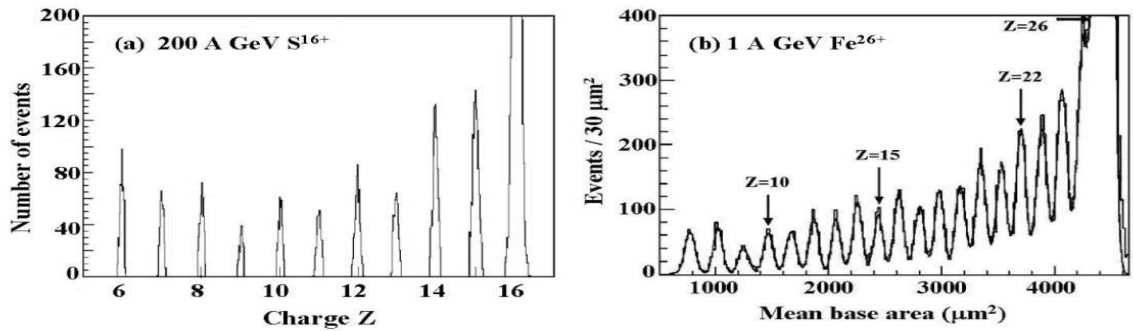}}\par}
\begin{quote}
\caption{\small (a) Charge distribution of 200 GeV/nucleon $S^{16+}$ ions and their fragments measured with CR39 NTDs (many measurements were made).
(b) Charge distribution measured with CR39 detector sheets located after a CH$_{2}$ target exposed to 1 A GeV Fe$^{26+}$ ions and their fragments.} 
\label{fig:illustration2}
\vspace{-0.5cm}
\end{quote}
%\end{center}
 \end{figure}

The fragmentation cross sections of various high energy nuclei on different nuclear targets were often measured using Nuclear Track Detectors (NTDs) \cite{ccec}. These measurements are of interest for nuclear physics and also for a number of applications, like cancer therapy, evaluation of the doses received by astronauts from cosmic rays \footnote{Cosmic rays are present everywhere in space: one can say that ``space is radioactive" because of the presence of cosmic rays. For future long range space explorations it is important to estabilish the fragmentation properties of heavy ions and the radioactive doses to astronauts.}, etc. Fig. \ref{fig:illustration2}a shows the charge distribution obtained with 200 GeV/nucleon S$^{16+}$ ions and their fragments produced in a copper target; notice the very good charge resolution for each peak obtained via the measurements of the nuclear fragments in 10 successive layers of CR39 NTDs [the resolution improves as the square root of the number of measurements]. Notice also the absence of nuclear fragments with fractional charge. One also observes the even/odd effect on the produced fragments (the height of a Z=even peak is higher than those of the close by peaks with Z=odd), seen in both fig. \ref{fig:illustration2}a,b. Fig. \ref{fig:illustration2}b shows the charge distribution of fragments from Fe$^{26+}$ ions of 1 GeV/nucleon in a CH$_{2}$ target; in this case only two layers were used for the measurement of the fragments; thus the charge resolution is not as good as in Fig. \ref{fig:illustration2}a.

\section{Experimentation at RHIC}
The Relativistic Heavy Ion Collider (RHIC) at Brookhaven lab (BNL) is capable of accelerating and colliding (in 2 identical rings) gold ions to c.m. energies of $\sqrt{s_{NN}}\sim$ 200 GeV/nucleon. The main purpose is to study the formation and the characteristics of the quark-gluon plasma, a state of matter believed to exist at sufficiently high energy densities \cite{ad}. If a RHIC collision produces a QGP, it will quickly cool, expand and coalesce into hadrons; thus experimental physicists cannot observe directly the QGP because its lifetime is too brief; they can study the hadrons that shower out of the collision. A collision that produces QGP will send out different kinds and ratios of particles than a collision that does not produce QGP. QGP is predicted by Quantum Chromodynamics; so theoretical physicists can calculate what signals it should produce. For example in the QGP jets are often produced, but some of them are suppressed by re-interactions in the dense QGP. In fact the QGP produced in a high energy nucleus-nucleus collision seems to behave more like a low viscosity liquid than a gas of free quarks and gluons \cite{ad}. 

\begin{figure}[!h]
\centering
\vspace{0.5cm}
{\centering\resizebox*{!}{7 cm}{\includegraphics{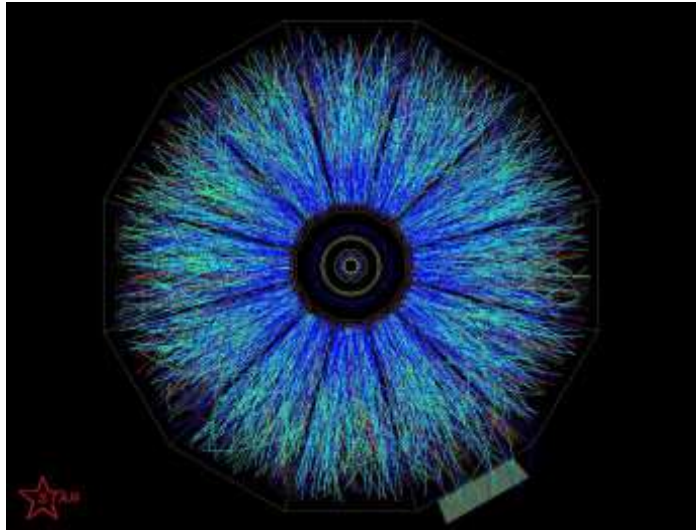}}}
\vspace{0.4cm}
\begin{quote} 
\caption{\small  One of the first events recorded by the STAR detector at RHIC in the collision of two gold ions at $\sqrt{s_{NN}} \sim$ 200 GeV/nucleon. The tracks indicate the paths of thousands of charged particles produced in the collision, as they pass through the STAR Time Projection Chamber.}
\label {fig:view}
\vspace{-0.5cm}
\end{quote}
\end{figure}

\begin{figure}[!t]
\centering
\vspace{-0.5cm}
{\centering\resizebox*{!}{8 cm}{\includegraphics
%[angle=180]
{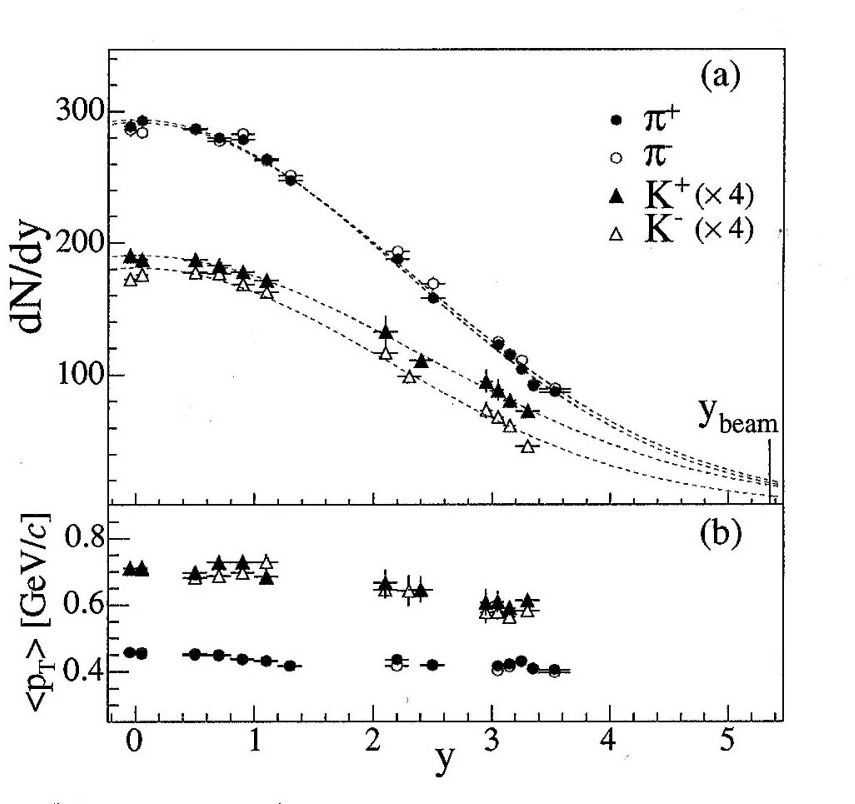}}}
%\vspace{0.4cm}
\begin{quote} 
\caption{\small (a) Pion and kaon rapidity densities, $dN/dy$, and (b) their mean transverse momentum $<p_{t}>$ as a function of rapidity $y$. The kaon yields were multiplied by a factor of 4 for clarity. The dashed lines in (a) are Gaussian fits to the $dN/dy$ distributions.}
\label {fig:fig7}
\end{quote}
\end{figure}

   The study of gold-gold high energy collisions is complicated also by the large number of charged particles produced, as seen in Fig. \ref{fig:view}, which shows one of the first events recorded by the STAR experiment at RHIC: in each collision about $\sim$4000 charged particles are produced, most of which are charged pions \cite{bnl}. As an example, Fig \ref{fig:fig7} shows the charged pion and kaon rapidity densities and the meam transverse momenta obtained by the BRAHMS collaboration at RHIC \cite{bea}. The number of  $\pi^{+}$ and $\pi^{-}$ are essentially equal, while the number of $K^{+}$ is slightly larger that the number of $K^{-}$. The average transverse momenta  decrease slowly with increasing rapidity y, for pions from 0.45 at $y=0$ to 0.40 at $y=3.5$; for kaons, from 0.71 at $y=0$ to 0.59 at $y=3.3$. The global number of $\pi^{+}$, $\pi^{-}$, $K^{+}$, $K^{-}$ is about 1660, 1683, 286, 242, respectively. 

   In addition to colliding heavy ions, RHIC is also able to collide deuteron-heavy ions and single protons ($pp$), at c.m. energies between 62 and 500 GeV/nucleon, and also polarized protons. The study of $pp$ collisions yields the high energy parameters, $\sigma_{tot}$, $\sigma_{el}$, slope of elastic scattering $B$, real to imaginary part, etc. The study of polarized proton collisions is interesting because it should provide further information on the proton spin. In the simplest quark model the proton spin is due to the spins of the three valence quarks, $u u d$. But one knows that dinamically the proton contains many gluons and sea quarks. Thus the explanation of the spin 1/2 of the proton must be more complex, and the study of high energy $pp$ collisions with polarized protons should provide further insight.
   
   As at any high energy accelerator, also particle search experiments are going on at RHIC.\\

\section{ Conclusions and outlook}
A wealth of experimental information was obtained since the 1960s on high energy $hh$ and h-nuclei collisions starting with simple beams and simple apparatus and then with better beams at higher energies and more complex apparatus. Most of the data were interpreted in terms of phenomenological models, which were eventually codified in Monte Carlo programs of increasing complexity \cite{mor}.
 
 Large area cosmic ray experiments should be able to improve the collision data in the ultra high energy region and solve some of the open problems, in particular the GZK cut off \cite{auger}.
 
 Also the amount of data on low $p_{t}$ inelastic processes is very large and also their interpretation was mainly performed with phenomenological models and Monte Carlo programs.
 Only the relatively small fraction of large $p_{t}$ events and of large $p_{t}$ jets of hadrons was interpreted in the context of perturbative QCD in terms of quarks and gluons.
 
 Hadron-nucleus collisions were often obtained as by-products; these data were often analyzed in terms of phenomenological models, using the Glauber theory, and complex Monte Carlos.
 
 The main enphasis in the study of high energy nucleus-nucleus collisions is connected with the search for the quark-gluon plasma, predicted by QCD. As stated in section 6, the situation is complicated by the fact that one cannot observe directly the QGP because its lifetime is too short: one can study the hadrons that shower out of the collision. Moreover the study is complicated by the very large number (thousands!) of charged hadrons produced; their detection requires complex and refined detectors with a very large number of electronies channels.
  
  There are some possible indications of QGP formation at the CERN SPS and at RHIC; the informations from RHIC are suggestive of a QGP which behaves more like a low viscosity liquid instead of a gas of free quarks and gluons.  
 
 The advent of LHC should allow exploration in a new energy region, which will be performed by large refined and complex general purpose detectors, refined detectors for the search for the QGP and smaller refined detectors for low $p_{t}$ physics \cite{link}.

  Professor Dumitru B. Ion contributed to important developments of quantum physics on various topics in the hadron and nuclear fields. Of particular interest are his papers on pionic \cite{ccce} and muonic radioactivity; searches and results  along these lines have been performed in many places around the world; a search is performed also at the Gran Sasso underground lab. \cite{ccce}. Professor Ion was awarded the Hurmuzescu Prize by the Romanian Academy of Sciences.

\vspace{1.5cm}
{\Large \bf Acknowledgements}
\\
\\
I am greatful to all the collaborators in the various experiments, total cross section and absorbtion measurements, beam survey measurements, measurements of low $p_{t}$ physics and of the general event structure, fragmentation measurements. I thank drs. Maddalena Errico, Roberto Giacomelli and Miriam Giorgini for technical support.

\bibliographystyle{plain}

\end{document}